\documentclass{article}
\usepackage{PRIMEarxiv}

\usepackage[utf8]{inputenc} 
\usepackage[T1]{fontenc}    
\usepackage{hyperref}       
\usepackage{url}            
\usepackage{booktabs}       
\usepackage{amsfonts}       
\usepackage{nicefrac}       
\usepackage{microtype}      
\usepackage{lipsum}
\usepackage{fancyhdr}       
\usepackage{graphicx}       
\usepackage{xcolor}
\usepackage[numbers]{natbib}

\title{Benefits and Limitations of Using GenAI for Political Education and Municipal Elections
}

\author{
  Raphael Fischer \\
  Lamarr Institute for ML and AI \\
  TU Dortmund University \\
  Dortmund, Germany\\
  \texttt{raphael.fischer@udo.edu} \\
   \And
  Youssef Abdelrahim \\
  TU Dortmund University \\
  Dortmund, Germany\\
  \texttt{youssef.abdelrahim@udo.edu} \\
   \And
  Katharina Poitz \\
  Lamarr Institute for ML and AI \\
  TU Dortmund University \\
  Dortmund, Germany\\
  \texttt{katharina.poitz@udo.edu} \\
}

\begin{document}
\maketitle

\begin{abstract}
Generative artificial intelligence (GenAI) presents both challenges and opportunities across all areas of education.
Facing the municipal elections in North Rhine-Westphalia, the \emph{Young AI Leaders} in Dortmund asked themselves---could GenAI be used to make political programs more accessible, in order to facilitate political education?
To explore respective potentials and limitations, we therefore performed an experimental study that combines different GenAI approaches.
Language models were used to automatically translate and analyze the contents of each program, deriving five potential visual appearance changes to the city of Dortmund.
Based on each analysis, we then generated images with diffusion models and published all results as an interactive webpage.
All GenAI models were locally deployed on a Dortmund-based computing cluster, allowing us to also investigate environmental impacts.
This manuscript explores the project in full depth, discussing technical details and critically reflecting on the results.
As part of the global Young AI Leaders Community, our work promotes the Sustainable Development Goal \emph{Quality Education} (SDG 4) by transparently discussing the pros and cons of using GenAI for education and political agendas.
\end{abstract}


\renewcommand{\arraystretch}{1.5}

\section{Introduction}

Modern generative artificial intelligence (GenAI) services are available to anyone with internet access, and thus play a critical role for all areas of education~\cite{tahiru_ai_2021}.
In the context of political activities, large language models (LLMs) can for example be utilized to build interactive voting advice applications~\cite{Schiele_2024} or to visualize political agendas~\cite{bundestagswahlai}. 
While this allows for interactive and customizable learning about politics, GenAI availability however also raises ethical questions with regard to misinformation, bias, and misuse~\cite{Saylam_Duman_Yildirim_Satsevich_2023}.
With regard to politics, LLMs could thus potentially be used for ``generative propaganda''~\cite{daepp2025generativepropaganda} and moreover have been analyzed to not be neutrally informative in itself~\cite{dormuth2025cautionarytaleneutrallyinformative,10817610}.
While these works already highlight potentials and limitations, additional empirical studies are needed to understand how AI can truly benefit political and ideological education.

\begin{figure}
    \begin{minipage}{.48\textwidth}
        \includegraphics[width=\linewidth]{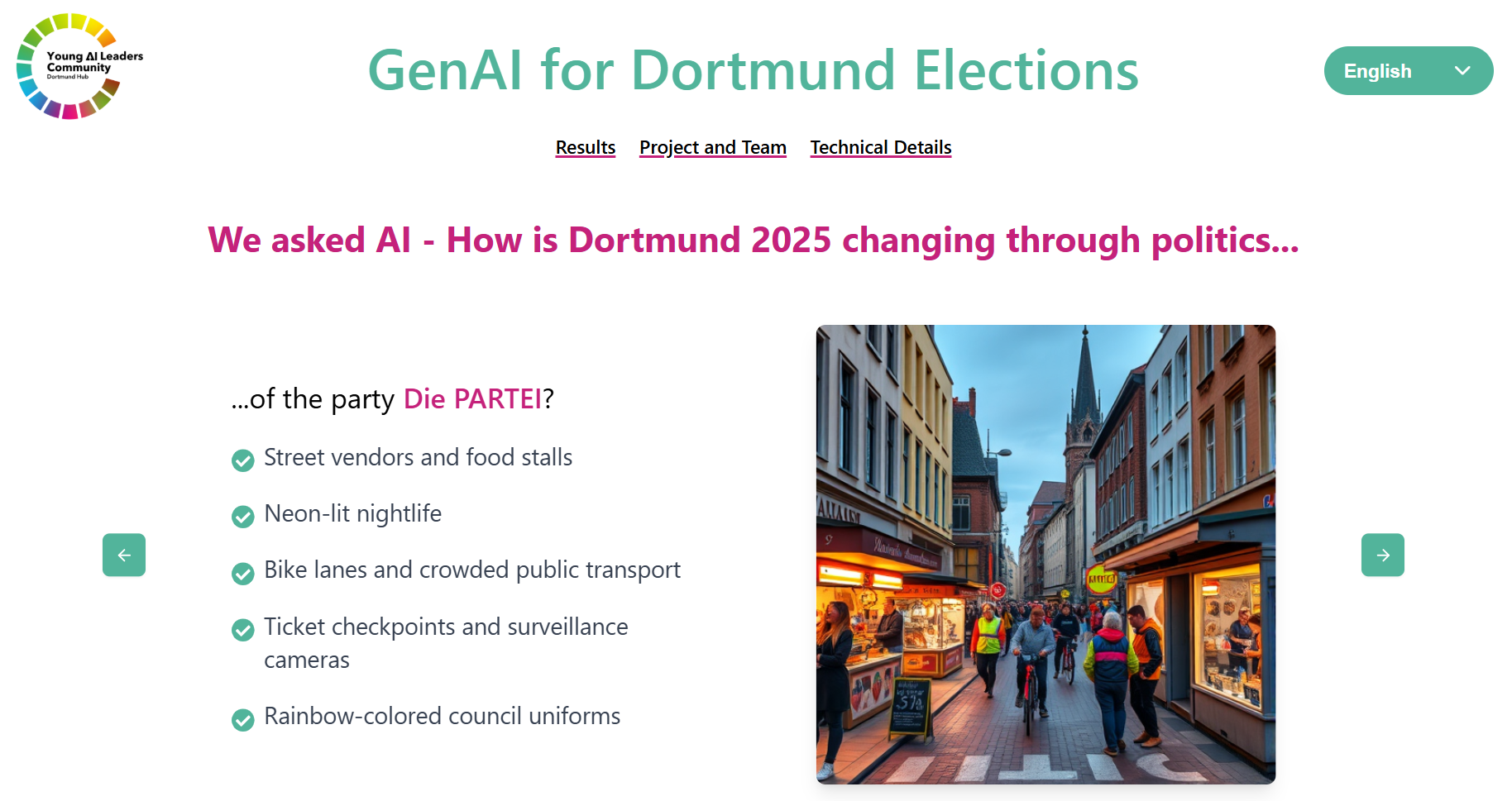} 
    \end{minipage}
    \hfill
    \begin{minipage}{.48\textwidth}
        \includegraphics[width=\linewidth]{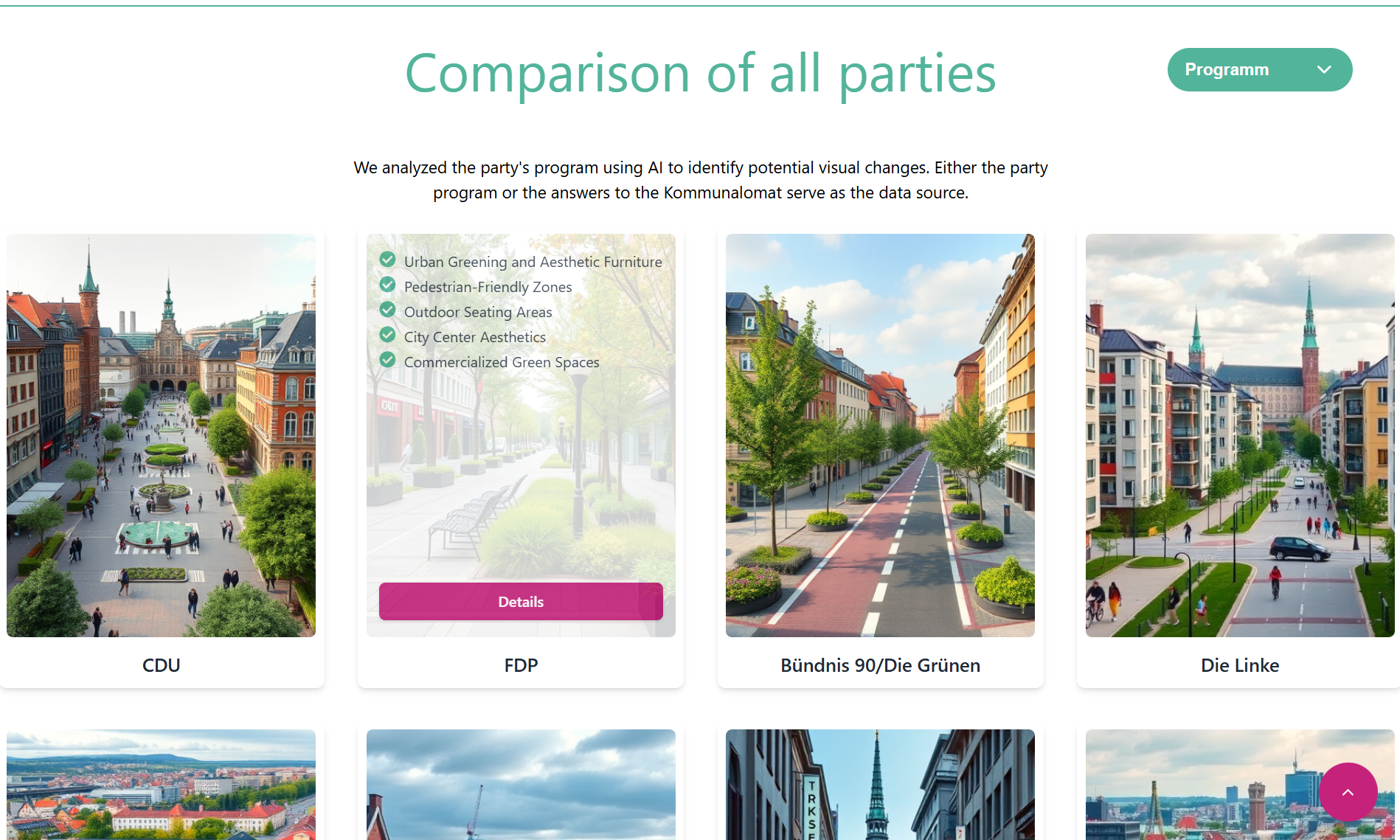} 
    \end{minipage}
    \caption{Website showcasing the \emph{Dortmund-Wahl-KI} results.}
    \label{fig:website}
\end{figure}

To further investigate the capabilities of GenAI for advancing political education, the Dortmund-based \emph{Young AI Leaders}\footnote{\url{https://youngaileaders-dortmund.de/}} have decided to launch the \emph{Dortmund-Wahl-KI} (Dortmund-Election-AI) project.
Our goal is to analyze and visualize local election programs with the help of GenAI, making the agendas more accessible in order to stimulate voter engagement and discourse.
For this project, we developed a custom AI pipeline and published all results in the form of an interactive webpage\footnote{\url{https://dortmund-wahl-ki.de/}}---see Figure~\ref{fig:website} for an overview. 
Our work is among the first projects of the worldwide \emph{Young AI Leaders} initiative and promotes the mission of \emph{AI for Good}\footnote{\url{https://aiforgood.itu.int/}}, who empower the global young to lead the AI revolution with positive change in mind.
In particular, the insights from this project shed light on the role of AI in the context of politics and education, thus relating to \emph{Quality Education} as the fourth Sustainable Development Goal (SDG 4)~\cite{boeren_understanding_2019}.
The manuscript at hand acts as a formal documentation of our project and is structured into four parts---we start with reviewing the literature related to \emph{Dortmund-Wahl-KI}, then explore the details of our methodology, after that present the published results, and lastly self-critically discuss the outcome of our project.
Our work follows open science practices, as we provide supplementary materials such as the experiment code and results in our public \emph{GitHub} repository~\footnote{\url{https://github.com/youngaileadersdortmund/dortmund-wahl-ki}}.
Moreover, we transparently report on the resource consumption of our AI pipeline, thus promoting the spirit of sustainable reporting~\cite{fischer_towards_2024,fischer_diss}.

\section{Related Work}

We first discuss existing literature that discusses how AI can be applied for political education and then discuss respective limitations and risks.

AI can be applied in many different educational areas~\cite{tahiru_ai_2021}, including but not limited to the context of ideologies and politics. 
Opportunities are for example seen for conforming to social advancement trends, meeting personalized learning demands, strengthening education efficiency, and facilitating the access to educational resources~\cite{10.1145/3660043.3660188}.
To give a specific first example, AI methods can be leveraged to analyze student behavior for providing personalized and interest-oriented learning content~\cite{10873211}.
AI can also be utilized through virtual reality or augmented reality, allowing for a brand new way of interactive learning~\cite{shan25opportunities}.
Similarly, GenAI models can be used to illustrate hypothetical or complex textual information, for example visualizing potential climate disasters~\cite{schmidt2022climategan} or political election programs~\cite{bundestagswahlai}, resulting in higher accessibility.
In the context of elections, conversational LLMs are moreover leveraged for offering interactive voting advice applications~\cite{Schiele_2024,dormuth2025cautionarytaleneutrallyinformative}.
Such initiatives result in an alternative representation of political agendas that is more accessible and inclusive than plain text documents, thus connecting to SDG 4 (\emph{Quality Education})~\cite{boeren_understanding_2019}.

With the educational benefits in mind, it is however also important to acknowledge associated risks.
The potential misuse of AI and LLMs in particular raises ethical concerns~\cite{Saylam_Duman_Yildirim_Satsevich_2023}, as they for example enable ``generative propaganda''~\cite{daepp2025generativepropaganda} as already actively adopted by German far-right actors~\cite{genai_far_right}.
In addition, GenAI is known to occasionally output factually wrong information, necessitating to establish ``hallucination awareness'' for students and teachers~\cite{Virtual_learning_genai}.
AI can also mirror and ``even amplify existing biases''~\cite{sci6010003}, for example due to unfair training data or biased algorithmic design choices.
Studies have for example evidenced that \emph{ChatGPT}, arguable the most prominent GenAI tool, exhibits biases relating to gender, ethnicity, religion, age, education, and social class~\cite{motoki_more_2024,qu_performance_2024}.
Research suggests that people tend to be more likely to accept and rely on AI when the bias aligns with user opinions~\cite{messer_how_2025}.
For the political domain, biases of AI are unfortunately more difficult to identify and eliminate~\cite{motoki_more_2024}, thus potentially having a strong influence on people~\cite{peters_algorithmic_2022}. 
Several studies have revealed that GenAI models commonly lean toward democratic and left-wing positions~\cite{dormuth2025cautionarytaleneutrallyinformative,ChatGPT_political_biases,socsci12030148,yuksel2025languagedependentpoliticalbiasai,motoki_more_2024}. 
Going into specifics, distinctions can be seen for different models:
Answering politically charged prompts, \emph{ChatGPT} and \emph{Claude} were found to exhibit a liberal bias, while \emph{Perplexity} takes more conservative stances and \emph{Gemini} has a centrist view~\cite{10817610}.
Another study suggests that \emph{Gemini} might actually be more left-wing and liberal-oriented than \emph{ChatGPT}~\cite{yuksel2025languagedependentpoliticalbiasai}, and moreover showed that the choice of query language can impact the resulting biases.
Naturally, understanding such biases is crucial when using GenAI, however a recent study also showcased that ``judgmental wording has a stronger impact on truthfulness predictions than political leaning'' in the context of automated fact checking~\cite{jakob2025polbixdetectingllmspolitical}.

\section{Methodology}

\begin{figure}
    \centering
    \includegraphics[width=\linewidth]{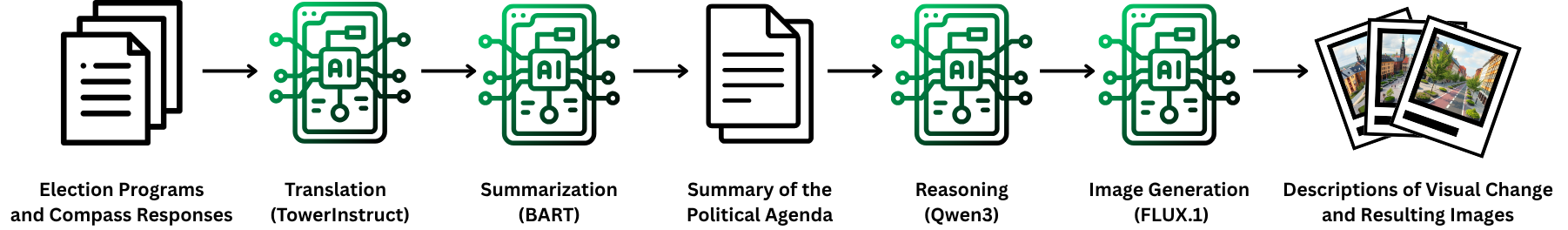}
    \caption{Schematic visualization of the GenAI pipeline behind \emph{Dortmund-Wahl-KI}.}
    \label{fig:pipeline}
\end{figure}

To further investigate the capabilities of GenAI for political education, we decided to perform an experimental study in the context of the 2025 municipal elections in Dortmund.
Our goal is to promote SDG 4 \emph{Quality Education} by raising awareness about benefits and limitations of using AI for making political content more accessible.
Similar to the \emph{bundestagswahl.ai} project~\cite{bundestagswahlai}, we used generative LLMs for analyzing and visualizing political agendas for the Dortmund election.
However, due to the aforementioned risks associated with GenAI, it was our goal to implement this project as transparently as possible. 
As such, instead of relying on black-box AI services like \emph{ChatGPT}, we implemented a specialized AI pipeline that combines different open-weights models and moreover allows for local (i.e., on-premise) deployment.
The different steps of the resulting analysis pipeline are summarized in Figure \ref{fig:pipeline} and will be explained in the following.

We started with collecting the election programs of different parties from their respective websites.
As a secondary data source, we also analyzed the individual party responses for the \emph{Kommunalomat} (local election compass)\footnote{\url{https://dortmund.waehlt.nrw/}}, obtained from the \emph{Jugendring (Youth Council) Dortmund}\footnote{\url{https://www.jrdo.de/}}.
In the first step of our pipeline, we enabled downstream processing with non-German language models by translating the primary data into English via \texttt{TowerInstruct-13B-v0.1}\footnote{\url{https://huggingface.co/Unbabel/TowerInstruct-13B-v0.1}}.
This model builds on an open-weights multilingual LLM, which was specialized for different tasks related to translation~\cite{tower_llm_2024}.
In our implementation, we split the text into chunks of ten sentences and fed them to the LLM with the following prompt:

\emph{Translate the following sentences from German into English: [sentences from the program]}

To reduce the complexity of the translated programs, we next compressed them with the help of \texttt{BART-Large-CNN}\footnote{\url{https://huggingface.co/facebook/bart-large-cnn}}, a fine-tuned denoising autoencoder for text summarization~\cite{DBLP:journals/corr/abs-1910-13461}.
These program summaries were then analyzed for potential visual changes of the city appearance.
The analysis was performed with the $\texttt{Qwen3-30B-A3B}$ reasoning model\footnote{\url{https://huggingface.co/Qwen/Qwen3-30B-A3B}}, which is a distilled variant of a powerful foundation model~\cite{yang2025qwen3technicalreport}. 
It uses a mixture-of-experts architecture for efficiency and also supports chain-of-thought inference, which results in more accurate and transparent outputs.
For analyzing the program summaries, we used the following prompt:

\emph{Identify five important visual aspects of a city appearance that would be affected or impacted by this political program. Describe each aspect in an informative and concise way, with 3 to 6 words. Return these five visual descriptions as a comma-separated list.}\\
\emph{[program summary]}

\begin{table}[t]
\centering
\caption{GenAI results based on party programs}
\begin{tabular}{p{5cm}p{5cm}p{5cm}}
\textbf{Christian Democratic Union}\footnote{\url{https://www.cdu-dortmund.de/}} & \textbf{Social Democratic Party}\footnote{\url{https://www.spd-dortmund.de/}} & \textbf{Alliance 90/The Greens}\footnote{\url{https://www.gruene-dortmund.de/}}\\
\hline
``Revitalized City Center, Lively Public Squares, Green Infrastructure, Safe Streets, Expanded Green Spaces'' & 
``Affordable housing developments, Modern school buildings, Expanded bike lanes, Barrier-free playgrounds, Green infrastructure expansion'' & 
``Increased urban greenery and tree coverage, expanded bike and pedestrian pathways, solar panel installations on buildings, de-paved and permeable surfaces, barrier-free street designs'' 
\\
\includegraphics[width=5cm]{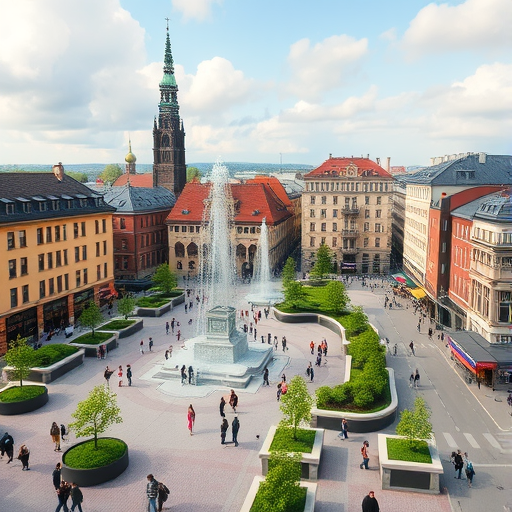} & \includegraphics[width=5cm]{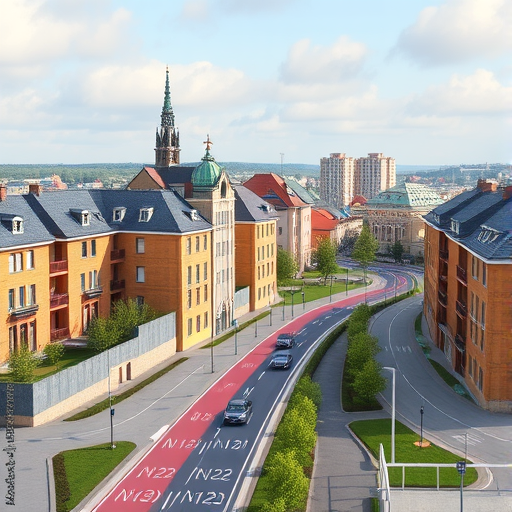} & \includegraphics[width=5cm]{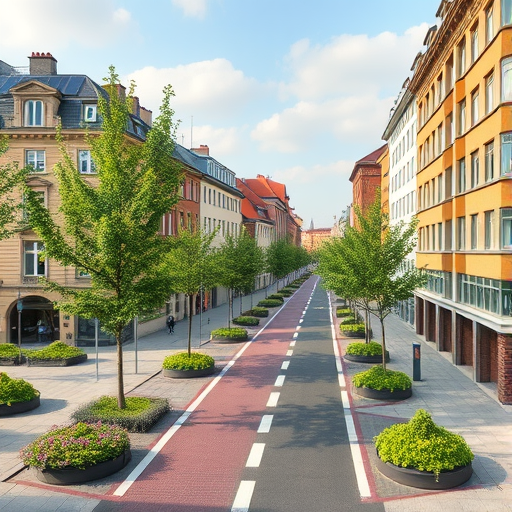} \\
\includegraphics[width=5cm]{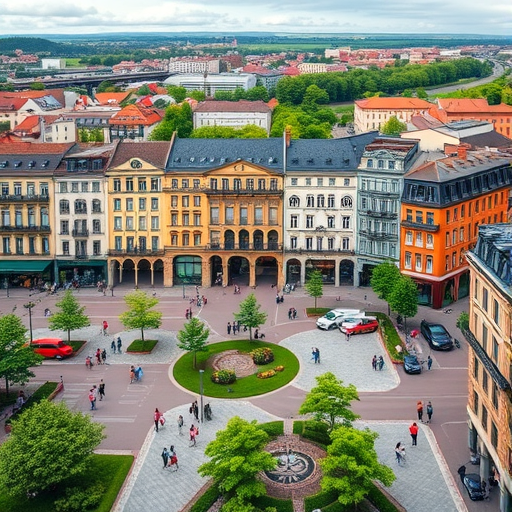} & \includegraphics[width=5cm]{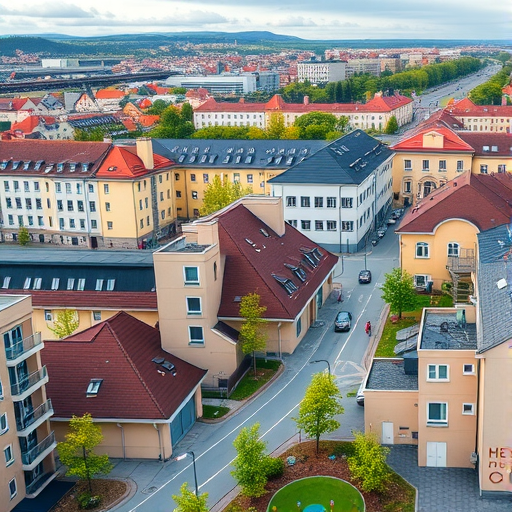} & \includegraphics[width=5cm]{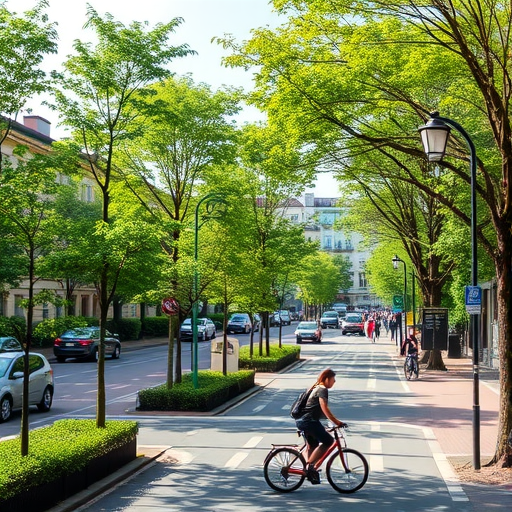} \\
\end{tabular}
\label{tab:results_program}
\end{table}

As the model provides visual descriptions and chain-of-thought output in English, we once more used $\texttt{TowerInstruct-13B-v0.1}$ to translate the reasoning output back to German, in order to better address the local audience.
Based on the English descriptions, we lastly used $\texttt{FLUX.1 [schnell]}$\footnote{\url{https://huggingface.co/black-forest-labs/FLUX.1-schnell}} for generating images that visualize the respective changes.
The base rectified flow transformer was trained using latent adversarial diffusion distillation and then optimized for fast and accurate image generation~\cite{labs2025flux1kontextflowmatching}.
In our pipeline, we used it with the following prompt, generating five image variants per query:

\emph{Dortmund city, with additional [visual descriptions]}

All AI experiments were performed on the computing cluster of the local Lamarr Institute\footnote{\url{https://lamarr-institute.org/}}, using a single $\texttt{NVIDIA A100}$ graphics processing unit (GPU). 
As already displayed in Figure~\ref{fig:website}, we presented the obtained results to the general public via an interactive webpage, powered by $\texttt{Vite}$\footnote{\url{https://vite.dev/}}, $\texttt{React}$\footnote{\url{https://react.dev/}}, $\texttt{JavaScript}$, and $\texttt{TailwindCSS}$\footnote{\url{https://tailwindcss.com/}}. 
The main page features a sliding element that showcases the analysis results for individual parties, as well as a grid of all images that provide detailed information upon user interaction (hover or click).
In addition, the website features pages for general information about the project and details on the technical realization.
We also placed a disclaimer on every page, clarifying the experimental nature of this project and pointing toward risks and biases of GenAI.
We offer German and English languages as well as desktop and mobile support, and deploy the website via $\texttt{GitHub Pages}$\footnote{\url{https://pages.github.com/}}.
To foster open science practices, we make the complete GenAI pipeline implementation, website code, and experimental results publicly available at \url{https://github.com/youngaileadersdortmund/dortmund-wahl-ki}.

\section{Results}

In total, we visualized eight party programs and corresponding election compass responses, resulting in $8 \cdot 2 \cdot 5 = 80$ images.
This section is limited to only discussing some selected results, however the rest can be explored online.

The GenAI outputs for three exemplary party programs are displayed in Table \ref{tab:results_program}, featuring the descriptions of potential visual changes obtained from AI reasoning as well as the respectively generated images (two variants each).
Comparing the textual results, we see a strong overlap---all programs were analyzed to advocate a green transformation, two of them mention expanded bike lanes, and they all seem to promote a modern and lively city.
Investigating the images below, some of these aspects seem to be well-represented, for example displaying urban greenery (all parties), a lively city center (CDU), and bike lanes (SPD and the Greens).
However, several other aspects cannot be found in the images---solar panels and modern buildings are missing, even though they would be easy to visualize, while characteristics such as safe streets, affordable housing, and permeable surfaces are arguably harder to incorporate in images.
In general, the images also seem to be quite similar in terms of daytime (bright daylight) and perspective (especially when comparing the variants per party).

\begin{table}[t]
\centering
\caption{GenAI results for three exemplary parties, based on election compass responses}
\begin{tabular}{p{5cm}p{5cm}p{5cm}}
\textbf{Christian Democratic Union} & \textbf{Social Democratic Party} & \textbf{Alliance 90/The Greens}\\
\hline
``Clean school exteriors, expanded public libraries, increased bike parking facilities, safer cycle paths with markings, more public shade trees'' & 
``Public Learning Spaces, Safer Cycle Paths, Bike Parking Spaces, Tree-lined Public Spaces, Solar Panel Installations'' & 
``Clean and well-maintained school areas, enhanced urban infrastructure, increased greenery and tree planting, improved public hygiene facilities, expanded pedestrian and cycling infrastructure '' 
\\
\includegraphics[width=5cm]{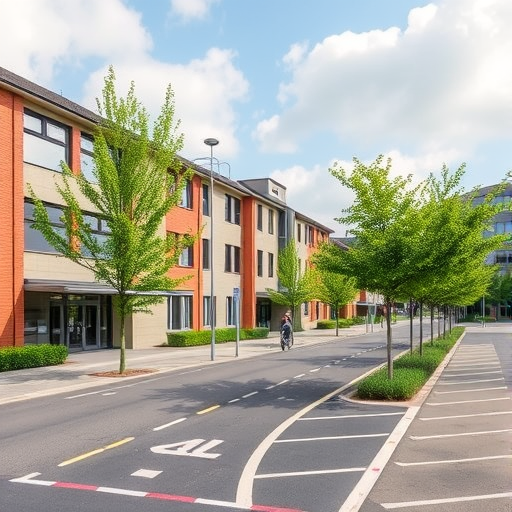} & \includegraphics[width=5cm]{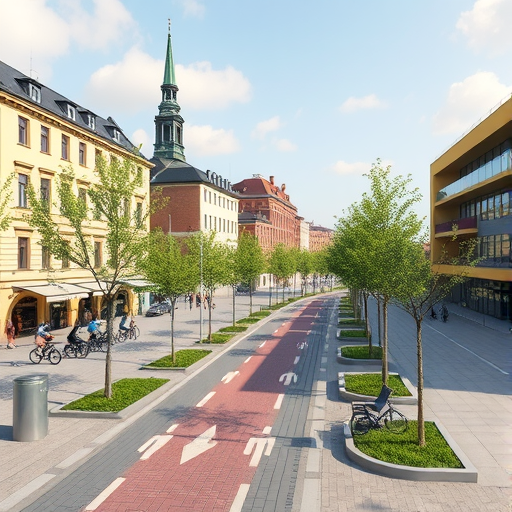} & \includegraphics[width=5cm]{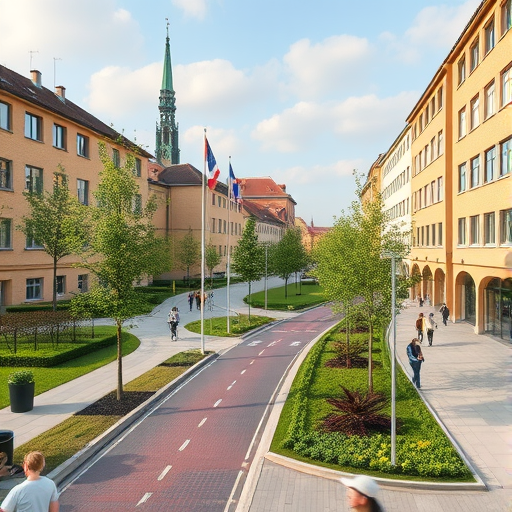} \\
\includegraphics[width=5cm]{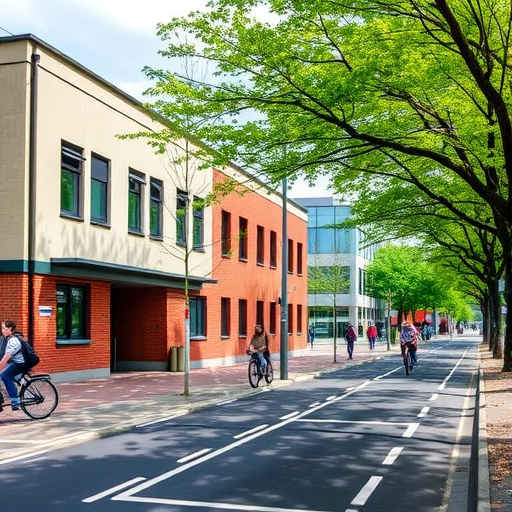} & \includegraphics[width=5cm]{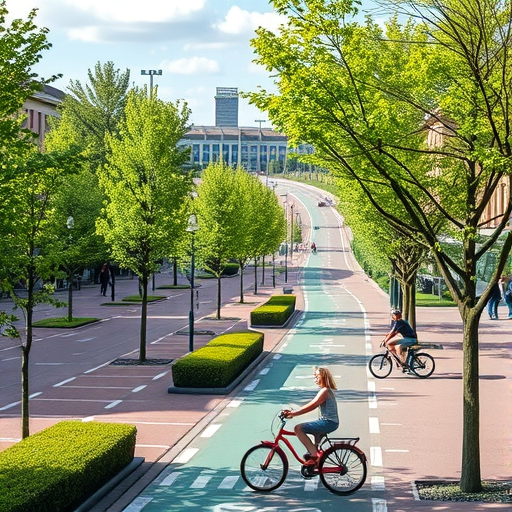} & \includegraphics[width=5cm]{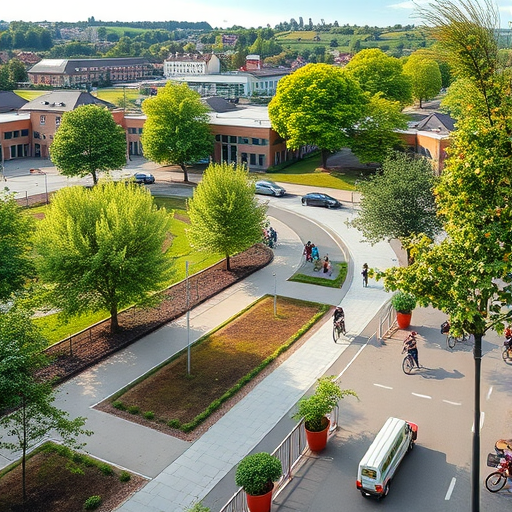} \\
\end{tabular}
\label{tab:results_kommunalomat}
\end{table}

In addition to the election programs, we used the responses of the parties for the local election compass (\emph{Kommunalomat}) as an alternative data source for our experiments.
These results are displayed in Table~\ref{tab:results_kommunalomat}, showcasing that the respectively analyzed visual points and resulting images are even more similar among the three selected parties.
For all of them, the AI analysis highlights clean school areas and cycling infrastructure, while more intricate details are hardly displayed in the images (libraries for CDU, solar panels for SPD, hygiene facilities for the Greens).
In these experiments, the nature of the primary data likely is a major issue---as the questions are the same for all parties, only \emph{approved} or \emph{rejected}, the AI analysis is likely biased toward the question wording itself, not the party response.
Recalling the reasoning prompt given in our Methodology, it is also interesting that both the program and election compass descriptions for the Greens is significantly longer than instructed, also in comparison to the other parties.
Another aspect of random behavior can be seen in the capitalization, which is not unified across the descriptions (despite using the same base prompt).

\begin{table}[t]
\centering
\caption{GenAI processing steps with corresponding duration, energy use, and emissions.}
\begin{tabular}{lcccc}
\hline
\textbf{Processing Step} & \textbf{Duration} & \textbf{Energy Use} & \textbf{Power Consumption} & \textbf{Emissions} \\
\hline
Translate (DE $\rightarrow$ EN) & 132.06 min & 0.71 kWh & 322.58 W & 0.27 kg CO\textsubscript{2}eq \\
Summarize        & 5.04 min   & 0.02 kWh & 238.10 W & 0.01 kgCO\textsubscript{2}eq \\
Reason           & 360.24 min & 1.41 kWh & 234.84 W  & 0.54 kg CO\textsubscript{2}eq \\
Translate (EN $\rightarrow$ DE) & 18.96 min  & 0.10 kWh & 316.46 W & 0.04 kg CO\textsubscript{2}eq \\
Generate Images  & 22.45 min  & 0.12 kWh & 320.71 W & 0.05 kg CO\textsubscript{2}eq \\
\hline
\textbf{Total}   & \textbf{538.75 min} & \textbf{2.37 kWh} & -- & \textbf{0.90 kg CO\textsubscript{2}eq} \\
\hline
\end{tabular}
\label{tab:processing_steps}
\end{table}

Promoting the spirit of sustainable reporting in AI~\cite{fischer_towards_2024,fischer_diss}, we also estimated the energy consumption and CO$_2$ emissions of running our AI pipeline.
Respective numbers were assessed with the help of \emph{CodeCarbon}~\cite{benoit_courty_2024_11171501}, as listed in Table \ref{tab:processing_steps}.
The GenAI models behind the individual processing steps vary in their complexity and resulting power demand.
The reasoning analysis had the highest environmental impact, mostly due to long processing times.
With these results, note that (1) \emph{CodeCarbon} is known to underestimate the actual energy demand of AI by 20--30\%~\cite{fischer2025groundtruthingaienergyconsumption} and (2) these results only reflect the environmental impact originating from the energy demand of one analysis run across all parties---as such, the reported numbers do not entail overhead impacts of project development, AI model life cycle aspects (e.g., training)~\cite{wu_sustainable_2022}, or embodied impacts of the utilized hardware~\cite{falk2025carboncradletograveenvironmentalimpacts}.

\section{Discussion}

We want to close our work with a self-critical discussion of our project, summarizing the results and linking them to established literature.
As stated in our Related Work section, we are aware that LLMs and GenAI models exhibit biases~\cite{dormuth2025cautionarytaleneutrallyinformative,ChatGPT_political_biases} and potentially do not correctly capture the political agendas~\cite{Virtual_learning_genai}.
At the same time, we face the fact that political actors already use GenAI for pushing their agendas~\cite{daepp2025generativepropaganda,genai_far_right} and that young people might include AI models in their (self-)education~\cite{Schiele_2024,bundestagswahlai}, despite the risks.
To us, this highlights the need for additional empirical studies and motivated us to conduct our project.
However, we strived for mitigating risks and maximizing transparency by following open science practices, using a reasoning model for the analysis, and prominently featuring a disclaimer on general problems of GenAI on the webpage.

Reflecting on our results, we believe that there are clear benefits but also strong limitations of using GenAI for political education.
Our project was positively received and sparked many interesting discussions around the municipal elections and political usage of AI, which otherwise would not have taken place.
On this personal scope, the feedback demonstrated how experimental projects can stimulate political education and discourse.
As a big benefit, GenAI capabilities for automated summarization and visualization can facilitate the exchange and learning about political agendas without being required to read extensive documents.
However, we also found limitations because (1) many aspects were not correctly represented in the images, (2) the visualizations generally look very similar, and (3) we found a strong gap when comparing the analysis outcomes of the programs and election compass answers.
A critical finding thus is the limited ability of our AI pipeline to capture subtle nuances and distinct priorities among different political agendas.
Instead, the results follow general trends that makes it hard to find differences among the actors.
Interestingly, the results of the closely related \emph{bundestagswahl.ai} project appear very different across all parties~\cite{bundestagswahlai}---unfortunately, due to the black-box nature of their utilized AI service (\emph{ChatGPT}), it is hard to understand why the results are so different.
One root cause for the diverging project outcomes might be that election programs are more similar on a municipal level than on the federal level, resulting in more similar AI visualizations.
Another option could be that \emph{ChatGPT} more explicitly focuses on (and potentially even exaggerates) the characteristics of the individual parties---it is unknown how much it actually pays attention to the given program, or just repeats biases from online reputation.
In short, our findings demonstrate that GenAI can serve political education, but it is extremely important to do so in a transparent, trustworthy, and responsible manner.

From these conclusions arise several opportunities for future research work.
On the technical site, one could aim at obtaining less similar results by developing an AI pipeline that focuses on the differences among political agendas, instead of conducting the analysis and visualization of each primary data source individually.
The significant limitations of the image generation model used in our work warrants further investigation and trying to find solutions that more accurately visualize the given aspects.
In addition to improving our proposed methodology with regard to  model selection and prompting, it could also applied to other political contexts and electoral analyses.
Interdisciplinary research could focus on evaluating human feedback on projects like ours in order to better characterize and potentially alleviate the limitations of GenAI for political education.
For the future, we also plan to collect and characterize the various opportunities for interdisciplinary work in the field via expert consulting.

\section{Conclusion}

The availability of GenAI necessitates to carefully investigate modeling capabilities in the context of education and politics.
Our project not only provided an overview of work in the field, but also is the first to analyze and visualize municipal election programs with LLMs and diffusion models.
The results highlight the potential for making political agendas more accessible, however also showcase the limitations for grasping intricate details and differences among parties.
With our work, we hope to make a helpful contribution to research on using AI for politics and education, while also promoting the sustainable use of AI, for good.

\bibliographystyle{unsrtnat}  
\bibliography{references}  

\end{document}